\documentclass[11pt]{article}
\usepackage{amssymb}

\newcommand{\BE}{\begin{equation}}
\newcommand{\EE}{\end{equation}}
\newcommand{\BA}{\begin{eqnarray}}
\newcommand{\EA}{\end{eqnarray}}

\begin{document}
\begin{titlepage}

\vspace*{1mm}
\begin{center}

            {\LARGE{\bf The Michelson-Morley experiment and \\ 
the cosmic velocity of the Earth }}

\vspace*{14mm}
{\Large  M. Consoli and E. Costanzo}
\vspace*{4mm}\\
{\large
Istituto Nazionale di Fisica Nucleare, Sezione di Catania \\
and Dipartimento di Fisica dell' Universit\`a \\
Via Santa Sofia 64, 95123 Catania, Italy}
\end{center}
\begin{center}
{\bf Abstract}
\end{center}
The Michelson-Morley experiment was designed to detect the 
relative motion of the Earth
with respect to a preferred reference frame, the ether, 
by measuring the fringe shifts in an optical interferometer. 
These shifts, that should have been proportional to the 
square of the Earth's velocity, were found to be much smaller than expected. 
As a consequence, that experiment
was taken as an evidence that there is no ether and, as such, 
played a crucial role for deciding between Lorentzian Relativity
and Einstein's Special Relativity. However, according to some authors,  
the observed Earth's velocity was {\it not} negligibly small. To provide an
independent check, we have re-analyzed
the fringe shifts observed in each of the six different sessions 
of the Michelson-Morley experiment. 
They are consistent with a non-zero {\it observable} Earth's velocity
$$ v_{\rm obs} = 8.4 \pm 0.5~ {\rm km/s}. $$
Assuming the existence of a preferred reference frame 
and using Lorentz transformations, this $v_{\rm obs}$ corresponds to a
{\it real} velocity, in the plane of the 
interferometer,
$$
v_{\rm earth} = 201 \pm 12~{\rm  km/s}.
$$
This value, which 
is  remarkably consistent with 1932 Miller's cosmic solution, suggests
that the magnitude of the 
fringe shifts is determined by the typical velocity of the
solar system within our galaxy. This conclusion is consistent with
the results of all classical experiments (Morley-Miller, 
Illingworth, Joos, Michelson-Pease-Pearson,...) and with
the existing data from present-day experiments. 
\vskip 35 pt
\end{titlepage}


{\bf 1.}~The Michelson-Morley experiment \cite{mm} is generally believed to 
represent the {\it proof} that the Earth's absolute motion
cannot be detected in a laboratory experiment. 
 However, the fringe shifts observed in the original experiment 
(and in the subsequent one of Morley and Miller \cite{morley}) 
although smaller than the expected magnitude corresponding to the orbital
motion of the Earth, were {\it not} negligibly small. 
While this had already been 
pointed out by Hicks \cite{hicks}, Miller's refined 
analysis of the half-period, 
second-harmonic effect observed in the experimental fringe shifts showed that
they were consistent with an
effective, {\it observable} velocity lying in the range 7-10 km/s
(see Fig.4 of Ref.\cite{miller}). 
For instance, the Michelson-Morley experiment gave a value
$v_{\rm obs} \sim 8.8 $ km/s for the noon observations 
and a value $v_{\rm obs} \sim 8.0 $ km/s for the evening observations.

The aim of this paper is twofold. On one hand, for the convenience of the
reader, we shall explicitly illustrate
some steps that are not immediately evident in the Michelson-Morley 
original paper and re-calculate the values of $v_{\rm obs}$ for their 
experiment.

On the other hand, by using Lorentz transformations, the small 
observed velocity will be shown to correspond to a {\it real} Earth's velocity, 
in the plane of the interferometer, 
$v_{\rm earth} \sim 200 $ km/s. This value, which is remarkably consistent
with 1932 Miller's cosmic solution \cite{miller}, suggests that the fringe
shifts are determined by the typical velocity of the solar system within our
galaxy (and not, for instance, by its velocity $v_{\rm earth} \sim 336$ km/s
with respect to the centroid of the Local Group). In this sense, 
this paper provides a consistent and self-contained treatment of the
Michelson-Morley type of experiments.
\vskip 10 pt

{\bf 2.}~We have analyzed 
the original data obtained by Michelson and Morley in each of
the six different sessions
of their experiment. No form of 
inter-session averaging has been performed. 
As discovered by Miller, in fact, 
inter-session averaging of the raw data may produce misleading results.
For instance, in the Morley-Miller data \cite{morley}, 
the morning and evening observations each were indicating
 an effective velocity of
about 7.5 km/s (see Fig.11 of Ref.\cite{miller}). This indication
 was completely lost
with the wrong averaging procedure adopted in Ref.\cite{morley}. 
The same point of view has been advocated by Munera in his recent 
re-analysis of the classical experiments \cite{munera}. 

To obtain the fringe shifts of each session we have
followed the well defined procedure adopted in the classical experiments as
described in Miller's paper \cite{miller}. Namely, starting from the seventeen
entries, say $E(i)$, reported in the Michelson-Morley Table \cite{mm}, 
one first has to correct for the difference $E(1) -E(17)$ 
between the first entry and the seventeenth entry
obtained after a complete rotation of the apparatus.
In this way, assuming the linearity of the
correction effect, one adds 15/16 of the correction
to the 16th entry, 14/16 to the
15th entry and so on, 
thus obtaining a set of 16 corrected entries 
\BE
E_{\rm corr}(i)={{i-1}\over{16}} (E(1)-E(17)) + E(i)
\EE
Finally, the fringe shift is defined from the 
differences between each of the corrected entries $E_{\rm corr}(i)$ and 
their average value $\langle E_{\rm corr} \rangle$ as
\BE
          \Delta \lambda (i)= E_{\rm corr}(i) - \langle E_{\rm corr} \rangle
\EE
We have fitted 
the amplitude $\bar{A}_2$ of the second-harmonic component
in a Fourier expansion ($\theta={{i-1}\over{16}} 2\pi$)
\BE
\label{fourier}
      {{\Delta \lambda(\theta)}\over{\lambda}} =\sum_n~\bar{A}_n 
\cos(n\theta+ \phi_n)
\EE
Following Miller's indications,  we have included terms up 
to $ n=5$, although 
the results for $\bar{A}_2$ are practically unchanged if one excludes from 
the fit the terms with $n=4$ and $n=5$. 
Our values of $\bar{A}_2$ for each session are reported in
Table 1. 

The Fourier analysis allows to 
determine the azimuth of the ether-drift effect, 
from the phase $\phi_2$ of the second-harmonic component, and an observable velocity 
from the value of its amplitude. 
To this end, we have used the basic relation of the experiment
\BE
                 2 \bar{A}_2= 
{{2D}\over{\lambda}} ~{{v^2_{\rm obs} }\over{c^2}}
\EE
where $D$ is the length of each arm of the interferometer.

Notice that, as emphasized by Shankland et al. (see page 178 of 
Ref.\cite{shankland}), it is the quantity 
$2 \bar{A}_2$, and not $\bar{A}_2$ itself, 
that should be compared with the maximal
displacement obtained for rotations of the apparatus through 
$90^o$ in its optical plane (see also Eqs.(\ref{fringe}) and (\ref{abar2})
below). Notice also that the quantity $2\bar{A}_2$ is denoted by $d$ in 
Miller's paper (see page 227 of Ref.\cite{miller}). 

 Therefore, 
for the Michelson-Morley apparatus where
 ${{D}\over{\lambda}}\sim 2\cdot 10^7$ \cite{mm}, 
 it becomes convenient to normalize the experimental values of
$\bar{A}_2$ to the classical prediction for an Earth's velocity of 30 
km/s 
\BE
{{2D}\over{\lambda}} ~{{(30 {\rm km/s})^2 }\over{c^2}}\sim 0.4
\EE
and we obtain
\BE
v_{\rm obs}  \sim 30 ~ \sqrt { 
{\bar{A}_2 }\over{ 0.2 } }~~{\rm km/s} 
\EE
Now, by inspection of Table 1, we find that
the average value of $\bar{A}_2$ from the noon sessions, 
$\bar{A}_2=0.017 \pm 0.003$, indicates a velocity
$v_{\rm obs}=8.7 \pm 0.8$ km/s
and the average value from the evening sessions, $\bar{A}_2=0.014 \pm 0.003$, 
indicates a velocity $v_{\rm obs}=8.0 \pm 0.8$ km/s. Since 
the two determinations
are well consistent with each other, we conclude that the 
Michelson-Morley experiment provides an observable velocity
\BE
\label{vobs}
       v_{\rm obs} = 8.4 \pm 0.5~{\rm km/s}
\EE
This 
is also in agreement with the results obtained by Miller himself at Mt. Wilson. 
 Differently from the original Michelson-Morley experiment
Miller's data were taken over the 
entire day and in four epochs of the year. 
However, after the critical re-analysis of Shankland et al. \cite{shankland}, 
it turns out that
the average daily determinations of $\bar{A}_2$ for the four epochs
are statistically consistent (see page 170 of Ref.\cite{shankland}).
 In this case, if one takes
the average of the four daily determinations, $\bar{A}_2=0.044 \pm 0.005$, 
one obtains a value which is just $\sim 1/13$ of 
the classical expectation for an Earth's velocity of 30 km/s
(see page 170 of Ref.\cite{shankland}) and an effective 
$v_{\rm obs}$ which is {\it exactly} the same as in Eq.(\ref{vobs}). 

The problem with Miller's analysis was to reconcile such low observable
 values of the Earth's velocity 
with those obtained from the daily {\it variations} of the magnitude and
azimuth of the ether-drift effect
with the sidereal time. In this way, in fact,  
on the base of the theory exposed by Nassau and Morse \cite{morse}, 
one can determine the apex of the motion of the
solar system. By requiring consistency among the four different
determinations obtained in the
four epochs of the year (see Fig.23 of Ref.\cite{miller}), Miller could
 restrict kinematically 
the cosmic Earth's velocity in the range 200-215
km/s (see page 233 of Ref.\cite{miller}) with the conclusion that 
"...a velocity $ v_{\rm earth} \sim 208 ~ {\rm km/s}$ 
for the cosmic component, gives the closest 
grouping of the four independently determined locations of the cosmic 
apex". 

At the same time, due to the
particular magnitude and direction of the cosmic component, Miller's 
predictions for its projection in the plane of the interferometer had very 
similar values (see Table V of Ref.\cite{miller}), say
\BE
\label{vcosmic}
v_{\rm earth} \sim 203 \pm 8 ~ {\rm km/s}
\EE
Therefore, after Miller's observations, the situation with the ether-drift 
experiments could be summarized as follows (see page 236 of Ref.\cite{miller}).
On one hand, "the observed displacement of the interference fringes, for some
unexplained reason, corresponds  to only a fraction of the velocity of the
Earth in space". On the other hand, the theoretical solution of the Earth's 
cosmic motion involves only the relative values of the ether-drift effect
and "..does not require a knowledge of the cause of the reduction in the
apparent velocity nor of the amount of this reduction". A check of this is 
that, after plugging the final parameters of the cosmic component in the
Nassau-Morse expressions, "..the calculated curves fit the observations 
remarkably well, considering the nature of the experiment" (see Figs. 26 and 27
of Ref.\cite{miller}). 

In spite of this beautiful agreement, 
the unexplained large discrepancy between the typical values of $v_{\rm obs}$, 
as given in Eq.(\ref{vobs}), and the typical calculated
values of $v_{\rm earth}$, as
given in Eq.(\ref{vcosmic}), has been representing a very
serious objection to the consistency of Miller's analysis. 

It has been recently 
pointed out, however, by Cahill and Kitto \cite{cahill} that an effective
reduction of the Earth's velocity from values
$v_{\rm earth}={\cal O}(10^2)$ km/s down to values
$v_{\rm obs}={\cal O}(1)$ km/s can be understood by
taking into
account the effects of the Lorentz contraction and of the
refractive index ${\cal N}_{\rm medium}$ of the
dielectric medium used in the interferometer. 

In this way, the
observations become consistent \cite{cahill} with values of the
Earth's velocity that are comparable to 
$v_{\rm earth} \sim 365$
km/s as extracted by fitting the COBE data for the cosmic
background radiation \cite{cobe}. The point is that the fringe shifts are
proportional to $ {{v^2_{\rm earth} }\over{c^2}}
 (1- {{1}\over{ {\cal N}^2_{\rm medium} }})$ rather than to
                  ${{v^2_{\rm earth} }\over{c^2}}$ itself. For the air, 
where ${\cal N}_{\rm air}\sim 1.00029$, assuming a value
$v_{\rm earth} \sim 365$ km/s, one would expect fringe shifts 
governed by an effective velocity 
$v_{\rm obs}\sim 8.8 $ km/s consistently with our value
Eq.(\ref{vobs}).

This would also explain why the experiments
of Illingworth \cite{illing} (performed in an apparatus filled with helium 
where ${\cal N}_{\rm helium}\sim 1.000036$) 
and Joos \cite{joos} (performed in the vacuum where
${\cal N}_{\rm vacuum}\sim 1.00000..$)
were showing smaller fringe shifts and, therefore, 
lower effective velocities.

In Ref.\cite{consoli} the argument has been completely reformulated by
 using Lorentz transformations (see also Ref.\cite{pagano}). 
As a matter of fact, in 
this case there is a non-trivial difference of a factor $\sqrt{3}$. 
When properly taken into account, the Earth's velocity extracted from
the absolute magnitude of the fringe shifts is {\it not} 
$v_{\rm earth} \sim 365$ km/s but $v_{\rm earth} \sim 201$ km/s thus
making Miller's prediction Eq.(\ref{vcosmic}) completely consistent with 
Eq.(\ref{vobs}). For the  convenience of the reader, we shall report in the
following the essential steps. 
\vskip 10 pt

{\bf 3.}~We shall start from
the idea that light propagates in 
 a medium with refractive index ${\cal N}_{\rm medium} > 1$  and
small Fresnel's drag coefficient
\BE
 k_{\rm medium}=
1- {{1}\over{ {\cal N}^2_{\rm medium} }} \ll 1
\EE
Let us
also introduce an isotropical speed of light 
($c=2.9979..10^{10}$ cm/s)
\BE
\label{u}
u\equiv  
{{c}\over{\cal{N}_{\rm medium} }}
\EE
The basic question is to determine experimentally, and
to a high degree of accuracy, 
whether light propagates
isotropically with velocity Eq.(\ref{u}) for an observer $S'$ placed on the
Earth. For instance for the air, where the relevant value is
${\cal N}_{\rm air}=1.00029..$, the isotropical value
${{c}\over{\cal{N}_{\rm air} }}$ is usually determined directly by
measuring the two-way speed of light along various directions. In this way, 
isotropy can be established at the level $\sim 10^{-7}$. 
If we require, however, 
a higher level of accuracy, say $10^{-9}$, the only way 
to test isotropy is to 
perform a Michelson-Morley type of experiment and look for 
fringe shifts upon rotation of the interferometer. 

Now, if one finds experimentally fringe shifts (and thus 
some non-zero anisotropy), 
one can explore the possibility that this effect is due 
to the Earth's motion with respect to a preferred frame 
$\Sigma\neq S'$. In this perspective, light would 
propagate isotropically with velocity as in Eq.(\ref{u})
for $\Sigma$ but {\it not} for $S'$. 

Assuming this scenario, the degree of anisotropy for $S'$ 
can easily be determined by using Lorentz
transformations. By defining
${\bf{v}}$ the velocity of
$S'$ with respect to $\Sigma$ one finds 
($\gamma= 1/\sqrt{ 1- {{ {\bf{v}}^2}\over{c^2}} }$) 
\BE
\label{uprime}
  {\bf{u}}'= {{  {\bf{u}} - \gamma {\bf{v}} + {\bf{v}}
(\gamma -1) {{ {\bf{v}}\cdot {\bf{u}} }\over{v^2}} }\over{ 
\gamma (1- {{ {\bf{v}}\cdot {\bf{u}} }\over{c^2}} ) }}
\EE
where $v=|{\bf{v}}|$. By keeping terms up 
to second order in $v/u$, one obtains
\BE
  {{ |{\bf{u'}}| }\over{u}}= 1- \alpha {{v}\over{u}} -\beta {{v^2}\over{u^2}}
\EE
where ($\theta$ denotes the angle between ${\bf{v}}$ and ${\bf{u}}$)
\BE
   \alpha = (1-  {{1}\over{ {\cal N}^2_{\rm medium} }} ) \cos \theta + 
{\cal O} ( ({\cal N}^2_{\rm medium}-1)^2 )
\EE
\BE
\beta = (1- {{1}\over{ {\cal N}^2_{\rm medium} }} ) P_2(\cos \theta) +
{\cal O} ( ({\cal N}^2_{\rm medium}-1)^2 )
\EE
with $P_2(\cos \theta) = {{1}\over{2}} (3 \cos^2\theta -1)$.

Finally defining $u'(\theta)= |{\bf{u'}}|$, the two-way speed of light is 
\BE
\label{twoway}
{{\bar{u}'(\theta)}\over{u}}= {{1}\over{u}}~ {{ 2  u'(\theta) u'(\pi + \theta) }\over{ 
u'(\theta) + u'(\pi + \theta) }}= 1- {{v^2}\over{c^2}} ( A + B \sin^2\theta) 
\EE
where 
\BE 
\label{ath}
   A= {\cal N}^2_{\rm medium} -1 + {\cal O} ( ({\cal N}^2_{\rm medium}-1)^2 )
\EE
and 
\BE
\label{BTH}
     B= -{{3}\over{2}} 
({\cal N}^2_{\rm medium} -1 )
+ {\cal O} ( ({\cal N}^2_{\rm medium}-1)^2 )
\EE
To address the theory of the Michelson-Morley
interferometer we shall consider
two light beams, say 1 and 2, that for simplicity are chosen 
perpendicular in $\Sigma$
where they propagate along the $x$ and $y$ axis with velocities
$u_x(1)=u_y(2)=
u= {{c}\over{ {\cal N}_{\rm medium} }}$.
Let us also assume that the velocity $v$ of  $S'$ is along the $x$ axis.

Let us now define $L'_P$ and $L'_Q$ to be the lengths of two
optical paths, say P and Q, as
measured in the $S'$ frame. For instance, they can represent the
lengths
of the arms of an interferometer which is at rest in the $S'$ frame.
In the first experimental set-up, the arm
of length $L'_P$ is taken along the direction of motion associated with the beam
1 while the arm of length $L'_Q$ lies along the direction of the beam 2.

In this way, the interference pattern, between the light beam coming out
of the optical path P and that coming out of the optical path Q, 
 can easily be obtained from the relevant delay time. 
 By using the
equivalent form of the Robertson-Mansouri-Sexl parametrization
\cite{robertson,mansouri} for the two-way speed of light 
defined above in Eq.(\ref{twoway}), this is given by
\BE
   \Delta T'(0) =
{{2L'_P}\over{\bar{u'}(0)}}- {{2 L'_Q}\over{\bar {u'}(\pi/2)}} 
\EE
On the other hand, if the beam 2 were to propagate along the optical path
P and
the beam 1 along Q, one would obtain a different
delay time, namely
\BE
   {(\Delta T')}_{\rm rot}=
{{2L'_P}\over{\bar{u'}(\pi/2)}}- {{2 L'_Q}\over{\bar {u'}(0)}} 
\EE
Therefore, by rotating the apparatus and
using Eqs.(\ref{ath}) and (\ref{BTH}), 
one obtains fringe shifts proportional to
\BE 
\label{deltat2} \Delta T'(0)- (\Delta T')_{\rm
rot} \sim (-2B) {{(L'_P+ L'_Q)}\over{u}}  
{{v^2}\over{u^2}} 
\EE 
or
\BE 
\label{deltat} 
\Delta T'(0)- (\Delta T')_{\rm
rot} \sim {{3(L'_P+ L'_Q)}\over{u}}  k_{\rm medium}
{{v^2}\over{u^2}} 
\EE 
(neglecting ${\cal O}(\kappa^2_{\rm medium})$ terms). 
This coincides with the pre-relativistic expression provided one replaces
$v$ with an effective observable velocity
\BE
\label{vobs0}
           v_{\rm obs}= v
\sqrt { k_{\rm medium} }
 \sqrt{3}
\EE
Finally, 
for the Michelson-Morley experiment, where $L'_P=L'_Q=D$, and for an ether wind
along the $x$ axis, 
the prediction for the fringe shifts at a given angle 
$\theta$ has the particularly simple form
\BE 
\label{fringe}
{{\Delta \lambda (\theta)}\over{\lambda}}=
  {{u \Delta T'(\theta) }\over{\lambda}} = {{u}\over{\lambda}} (
{{2D }\over{\bar{u'}(\theta)}}- {{2 D }\over{\bar {u'}(\pi/2+\theta)}})=
 {{2 D }\over{\lambda}} {{v^2}\over{c^2}} (-B) \cos (2\theta) 
\EE
that corresponds to a pure second-harmonic 
effect. At the same time, it becomes clear the remark by 
Shankland et al. (see page 178 of Ref.\cite{shankland})
that its amplitude 
\BE
\label{abar2}
              \bar{A}_2 \equiv  
{{2 D }\over{\lambda}} {{v^2}\over{c^2}} (-B) = 
{{D }\over{\lambda}} {{v^2_{\rm obs} }\over{c^2}}  
\EE
is just one-half of the corresponding quantity entering Eq.(\ref{deltat2}). 

\vskip 10 pt
{\bf 4.}~Now, if upon operation of the interferometer
there are fringe shifts  and if their magnitude,
 observed with different dielectric media and within the
experimental errors, points
consistently to a unique value of the Earth's velocity, there is
experimental evidence for
the existence of a preferred frame $\Sigma \neq S'$. In practice, to
 ${\cal O}({{v^2_{\rm earth} }\over{c^2}} )$, this can be decided by
re-analyzing \cite{cahill}
the experiments in terms of the effective parameter
 $\epsilon = {{v^2_{\rm earth} }\over{u^2}} k_{\rm medium}$. The
conclusion of Cahill and Kitto \cite{cahill} 
is that the classical experiments are consistent with the value
 $v_{\rm earth}\sim 365$ km/s obtained from the COBE data. 

However, in
our expression Eq.(\ref{vobs0}) determining the fringe shifts there is a difference
of a factor $\sqrt{3}$ with respect to their result
$v_{\rm obs}=v \sqrt { k_{\rm medium} }$. Therefore, using
Eqs.(\ref{vobs0}) and (\ref{vobs}), for ${\cal N}_{\rm air} \sim 1.00029$, 
 the relevant Earth's velocity (in the plane of the interferometer)
 is {\it not} $v_{\rm earth}\sim 365$ km/s but rather
\BE
\label{vearth}
                  v_{\rm earth} \sim 201 \pm 12 ~{\rm km/s}
\EE
in excellent agreement with the value Eq.(\ref{vcosmic}) calculated by Miller. 

Therefore, from this excellent agreement, we deduce that
the magnitude of the fringe shifts is determined by the typical
velocity of the solar system within our galaxy and not, for instance, 
by its velocity relatively to the
centroid of the Local Group. 
In the latter case, one would get higher values such as
$v_{\rm earth} \sim 336$ km/sec, see Ref.\cite{V5}. 

Notice that such ambiguity, say  
$v_{\rm earth}\sim 200, 300, 365,...$ km/s, on the actual value of the
Earth's velocity determining
the fringe shifts, can only be resolved experimentally in view of the
many theoretical uncertainties in the operative
definition of the preferred frame
where light propagates isotropically. At this stage, we believe, one should 
just concentrate on the internal consistency of the various frameworks. 
In this sense, the analysis presented in this paper shows
that internal consistency is extremely high in Miller's 1932 solution. 

We are aware that our conclusion goes against the widely spread belief 
that Miller's results were only due to statistical fluctuation and/or 
local temperature conditions (see the Abstract of Ref.\cite{shankland}). 
However, within the paper the same authors of Ref.\cite{shankland}
say that "...there can be 
little doubt that statistical fluctuations alone  cannot account for the
periodic fringe shifts observed by Miller" 
(see page 171 of Ref.\cite{shankland}).  In fact, although "...there is 
obviously considerable scatter in the data at each azimuth position,...the
average values...show a marked second harmonic effect"
(see page 171 of Ref.\cite{shankland}). In any case, interpreting the observed
effects on the base of the local temperature conditions cannot be the 
whole story since "...we must admit that a direct and general 
quantitative correlation between amplitude and phase of the observed 
second harmonic on the one hand and the thermal conditions in the observation
hut on the other hand could not be established" 
(see page 175 of Ref.\cite{shankland}). This rather
unsatisfactory explanation of the observed effects
should be compared 
with the previously mentioned
excellent agreement that was instead obtained by Miller
once the final parameters for the Earth's velocity were plugged in the 
theoretical predictions (see Figs.26 and 27 of Ref.\cite{miller}).

This does not exclude the presence of some systematic effect in the
Miller's data. In fact, as mentioned above, Miller's value
$\bar{A}_2=0.044 \pm 0.005$, that perfectly agrees with Eq.(\ref{vobs}),
 was only obtained after the critical 
re-analysis of Shankland et al. (see page 170 of Ref.\cite{shankland}). 

On the other hand, 
additional information on the validity of the Miller's results
can also be obtained by other means, for instance 
comparing with the experiment performed by Michelson, 
Pease and Pearson \cite{mpp}. These other authors in 1929, 
using their own interferometer, again at Mt. Wilson, declared that 
their "precautions taken to eliminate effects of 
temperature and flexure disturbances were effective". Therefore, their statement that the
fringe shift, 
as derived from "..the displacements observed at maximum and minimum at 
sidereal times..", was definitely smaller than "...one-fifteenth of
that expected on the 
supposition of an effect due to a motion of the solar system of three 
hundred kilometres per second", can be taken as an indirect 
confirmation of Miller's results. Indeed,
although the "one-fifteenth" was actually 
a "one-fiftieth" (see pag.240 of Ref.\cite{miller}), 
their fringe shifts were certainly non negligible. This is easily understood 
since, for an in-air-operating interferometer, 
the fringe shift $(\Delta\lambda)_{\rm class}(300)$, expected on the base of 
classical physics
for an Earth's velocity of 300 km/s, is about 500 times
bigger than the corresponding relativistic one
\BE
(\Delta\lambda)_{\rm rel}(300)\equiv 3 k_{\rm air}
~ (\Delta\lambda)_{\rm class}(300)
\EE
computed using Lorentz transformations 
(compare with Eqs.(\ref{deltat}) and (\ref{vobs0}) for 
$k_{\rm air}\sim {\cal N}^2_{\rm air} -1 \sim 0.00058$). 
 Therefore, the Michelson-Pease-Pearson upper bound
\BE
(\Delta\lambda)_{\rm obs}< 0.02~
 (\Delta\lambda)_{\rm class} (300)
\EE
is actually equivalent to
\BE
(\Delta\lambda)_{\rm obs}< 24 ~
 (\Delta\lambda)_{\rm rel} (208)
\EE
As such, it poses no strong restrictions and is entirely 
consistent with those typical low effective velocities detected
by Miller in his observations of 1925-1926. 

A similar agreement is obtained when comparing with the Illingworth's data
\cite{illing} as recently 
re-analyzed by Munera \cite{munera}. In this case, using Eq.(\ref{vobs0}), 
from
the observable velocity $v_{\rm obs}=3.13 \pm 1.04$ km/s \cite{munera} 
and the value
$N_{\rm helium} \sim 1.000036$, 
 one deduces $v_{\rm earth}=213 \pm 71$ km/s, in very good agreement with 
Eq.(\ref{vcosmic}). 

The same conclusion applies to the Joos experiment 
\cite{joos}. His interferometer was placed in an evacuated housing and
he declared that the velocity of any ether wind 
had to be smaller than 1.5 km/s.
Although we don't know the exact 
value of $N_{\rm vacuum}$ for the Joos experiment, 
it is clear that this is the type of upper bound
expected in this case. As an example, for $v_{\rm earth}\sim 208$ km/s, one
obtains $v_{\rm obs}\sim 1.5$ km/s for 
$N_{\rm vacuum}\sim 1.000009$ and 
$v_{\rm obs}\sim 0.5$ km/s for 
$N_{\rm vacuum}\sim 1.000001$. In this sense, the effect of using Lorentz 
transformations is most 
dramatic for the Joos experiment when comparing with
the classical expectation for an Earth's velocity
of 30 km/s. Although the relevant Earth's velocity 
is $\sim 208$ km/s, 
the fringe shifts, rather than being $\sim 50$ times {\it bigger} than the 
classical prediction, are  $\sim 1000$ times {\it smaller}. 

\vskip 10 pt

{\bf 5.}~We shall conclude with a brief comparison with present-day, 
 `vacuum' Michelson-Morley experiments of the type first performed by 
Brillet and Hall \cite{brillet} and more recently by 
M\"uller et al. \cite{muller}. In a perfect vacuum, by definition 
${\cal N}_{\rm vacuum}=1$ so that $v_{\rm obs}=0$ and no anisotropy can 
be detected. However, one can explore 
\cite{pagano,modern} the possibility that, even in this case,
 a very small anisotropy might be due to a refractive index 
${\cal N}_{\rm vacuum}$ that differs from unity by an infinitesimal
amount. In this case, the natural candidate to explain a value
${\cal N}_{\rm vacuum} \neq 1$
is gravity. In fact, by using
the Equivalence Principle, any freely falling frame $S'$ will locally 
measure the same speed of light as in an inertial frame in the absence of
any gravitational effects. However, if $S'$ carries on board an heavy 
object this is no longer true. For an observer placed on the Earth, 
this amounts to insert
the Earth's gravitational potential in the  weak-field isotropic
approximation to the line element of
 General Relativity \cite{weinberg}
\BE
ds^2= (1+ 2\varphi) dt^2 - (1-2\varphi)(dx^2 +dy^2 +dz^2)
\EE
so that one obtains a refractive index for
light propagation 
\BE
\label{nphi}
            {\cal N}_{\rm vacuum}\sim  1- 2\varphi
\EE
This represents the `vacuum' analogue of 
${\cal N}_{\rm air}$, ${\cal N}_{\rm helium}$,...so that from
\BE 
     \varphi =- {{G_N M_{\rm earth}}\over{c^2 R_{\rm earth} }} \sim
-0.7\cdot 10^{-9}
\EE
and using Eq.(\ref{BTH}) one predicts
\BE
\label{theor}
                 B_{\rm vacuum} \sim -4.2 \cdot 10^{-9}
\EE
For $v_{\rm earth} \sim 208$ km/s, 
this implies an observable anisotropy of the two-way speed of light 
in the vacuum Eq.(\ref{twoway}) 
\BE
        {{ \Delta \bar{c}_\theta }\over{c}} \sim 
|B_{\rm vacuum}| {{v^2_{\rm earth} }\over{c^2}} \sim 2\cdot 10^{-15}
\EE
in good agreement with the experimental value
        ${{ \Delta \bar{c}_\theta }\over{c}}= (2.6 \pm 1.7) \cdot 10^{-15}$ 
determined by M\"ueller et al.\cite{muller}.

\vfill
\eject
\begin{table}
\centering{\begin{tabular}{c|c}  
SESSION       &       $\bar{A}_2$   \\
July 8  (noon) & $0.010 \pm 0.005$  \\ 
July 9  (noon) & $0.015 \pm 0.005$   \\
July 11 (noon) & $0.025 \pm 0.005$    \\
July 8  (evening) & $0.014 \pm 0.005$  \\
July 9  (evening) &$0.011 \pm 0.005$   \\
July 12 (evening) & $0.018 \pm 0.005$  \\ 
\end{tabular}
\vskip 20 pt
\caption{\rm
We report the amplitude of the second-harmonic component
$\bar{A}_2$ obtained from the fit Eq.(\ref{fourier}) 
to the various samples of data.}
\label{tab:1}}
\end{table}

\end{document}